\documentclass{article}
\usepackage{epsfig,amsmath}
\begin{document}
\begin{flushright}
hep-ph/0310353 \\
UG-FT-158/03, CAFPE-29/03, DFPD-03/TH/41 \\
October 2003\\
\end{flushright}
\vspace*{5mm}
\begin{center}

{\Large\bf Physics of Brane Kinetic
  Terms\footnotemark[1]\footnotetext[1]{Presented at Matter To The Deepest:
Recent Developments In Physics of Fundamental Interactions,
XXVII International Conference of Theoretical Physics,
Ustron 15-21 September 2003, Poland.}}

\vspace{1.4cm}
{\sc F. del Aguila$^1$, M. P\'erez-Victoria$^2$ and J. Santiago$^3$}\\
\vspace{.5cm}
{\it $^1$Centro Andaluz de
F\'\i sica de Part\'\i culas Elementales (CAFPE)}\\
{\it and Departamento de F\'\i sica Te\'orica y del Cosmos}\\
{\it Universidad de Granada, E-18071 Granada, Spain}\\
\vspace{.5cm}
{\it $^2$Dipartimento di Fisica ``G. Galilei'', Universit\`a di
  Padova and}\\
{\it INFN, Sezione di Padova, Via Marzolo 8, I-35131 Padua, Italy}\\
\vspace{.5cm}
{\it $^3$IPPP, Centre for Particle Theory}\\
{\it University of Durham}\\
{\it DH1 3LE, Durham, U.K.}\\

\end{center}
\vspace{1.cm}
\begin{abstract}

Models with extra dimensions may give new effects visible at future
experiments. In these models, bulk fields can develop localized
corrections to their kinetic terms which can modify the
phenomenological predictions in a sizeable way. We review the case in
which  both gauge bosons and fermions propagate in the bulk,
and discuss the limits on the parameter space arising from electroweak
precision data.  

\end{abstract}

%%%%%%%%%%%%%%%%%%%%%%%%%%%%%%%%%%%%%%%%%%%%%%%%%%
%                                                %
%    BEGINNING OF TEXT                           %
%                                                %
%%%%%%%%%%%%%%%%%%%%%%%%%%%%%%%%%%%%%%%%%%%%%%%%%%
%\begin{document}
%\title{Physics of Brane Kinetic Terms
%\thanks{Presented at ...}%
%\author{F. del Aguila,
%\address{Departamento de F{\'\i}sica Te\'orica y del Cosmos and
%Centro Andaluz de F{\'\i}sica de Part{\'\i}culas Elementales
%(CAFPE), Universidad de Granada, E-18071 Granada, Spain}
%%\and
%\\[.25cm]
%M. P\'erez-Victoria
%\address{Dipartimento di Fisica ``G. Galilei'', Universit\`a di Padova
%and INFN, Sezione di Padova, Via Marzolo 8, I-35131 Padua, Italy}
%\and
%J. Santiago
%\address{Institute for Particle Physics Phenomenology, University of Durham,
%South Road, Durham DH1 3LE, UK}
%}
%\maketitle
%\begin{abstract}
%Models with extra dimensions may give new effects visible at future
%experiments. In these models, bulk fields can develop localized
%corrections to their kinetic terms which can modify the
%phenomenological predictions in a sizeable way. We review the case in
%which  both gauge bosons and fermions propagate in the bulk,
%and discuss the limits on the parameter space arising from electroweak
%precision data.  
%\end{abstract}
%\PACS{11.10.Kk, 11.25.Wx, 11.90.+t, 12.90.+b}

\section{Introduction}

Models with extra dimensions have received a lot of attention 
in the last few years (see \cite{Hewett} for a review of their 
phenomenology). Besides being motivated by string 
theory, they provide new mechanisms to face
longstanding problems, such as the Planck to electroweak and
electroweak to cosmological constant hierarchy problems, the fermion
flavour problem, symmetry and supersymmetry breaking among others. 
The original brane world idea in  
which only gravity propagates in the extra dimensions and matter and 
gauge fields are bounded to a four-dimensional brane was soon extended 
to allow for gauge fields and even all matter 
fields propagating in the bulk of extra dimensions. Even in this case, 
lower dimensional submanifolds (branes, domain walls, orbifold fixed 
points or planes, \ldots) are usually present in the models. In 
Ref.~\cite{Dvali00} it was argued that in the original brane world 
model, matter and gauge loop corrections to the graviton self-energy
generate a curvature term in the action localized at the
position of the brane, which in turn has stricking
phenomenological implications, with brane scientists observing
four-dimensional gravity up to 
cosmological distances even in the case of an infinite flat 
bulk. Even if this particular model has strong coupling problems
problems~\cite{Luty03}, the idea of Brane 
localized Kinetic Terms (BKT) may have very important phenomenological
consequences not only for gravitational physics but for
any kind of bulk field. As a matter of fact, these terms are 
unavoidable in interacting theories with extra dimensions and
lower-dimensional defects. The reason is that translational invariance
is broken by the defects, and this allows for localized
divergent radiative corrections, which must be cancelled by the
corresponding localized counterterms. In particular, BKT are induced
in this way. Their coefficients run with the scale, so they cannot be
set to zero at all energies. This suggests that one should include
them already at tree level, which means that they are not necessarily
loop suppressed. The generation of BKT by radiative corrections in
orbifolds without brane couplings at tree level was explicitly shown
in a particular example in~\cite{Georgi01}.

A detailed study of the implications of general BKT for bulk fields of spin
0, 1/2 and 1 has been carried out in Ref.~\cite{Aguila03a}. There, it
was shown that some of the BKT  
one can write have a smooth behaviour as they get small, whereas
others give rise to a singular behaviour in the spectrum and have to
be dealt with using classical renormalization order by order in 
perturbation theory. In the following, we discuss in a pedagogical
manner the properties and
phenomenology of fermions~\cite{Aguila03a,Aguila03b} 
and gauge bosons~\cite{Carena02,Carena03}, studying
in some detail the case in which both have BKT. (For a
compendious review of BKT with a more complete list of references,
see~\cite{Aguila03c}.) 
For simplicity, we will focus on BKT which do not need classical
renormalization.  

%%%%%%%%%%%%%%%%%%%%%%%%%%%%%%%%%%%%%%%%%%%%%%%%%%%
 
\section{Brane kinetic terms for bulk fermions and gauge bosons}
\label{sectionKK} 
We consider a five-dimensional model with the fifth dimension
$y$ compactified on an orbifold $S^1/Z_2$, that is to say, 
a circle of radius $R$ with opposite points identified: $y 
\sim -y$. The four dimensional
subspaces $y=0,\pi R$ are fixed under 
the $Z_2$ action. From now on we will call these fixed hyperplanes
``branes'', although they are static non-fluctuating objects. As argued
in the introduction, the action contains in general kinetic terms
localized on these branes, besides the usual Poincar\'e invariant ones.

\begin{figure}[!t]
\begin{center}
\epsfig{file=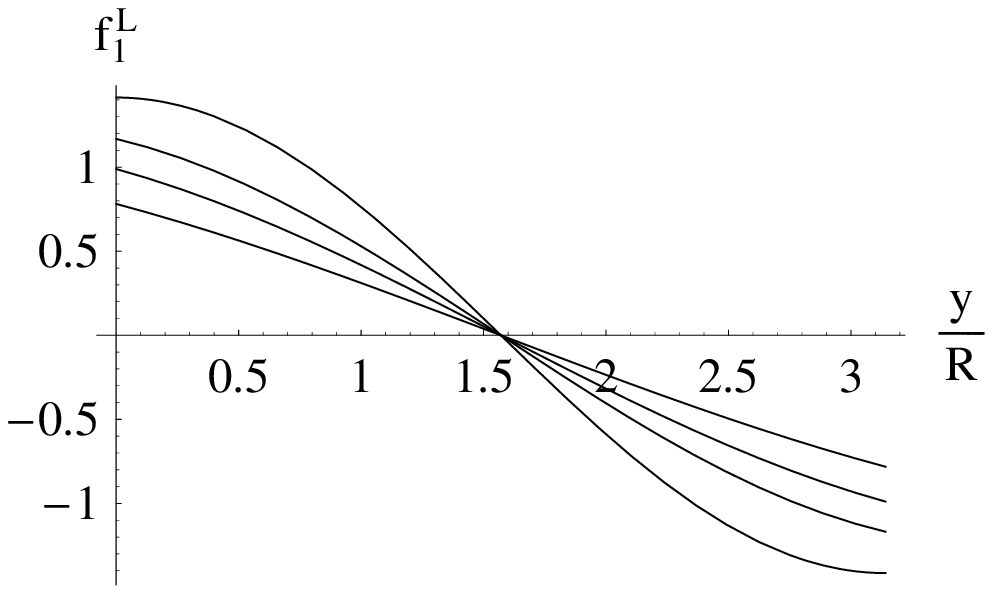,width=5cm}\hspace{.5cm}
\epsfig{file=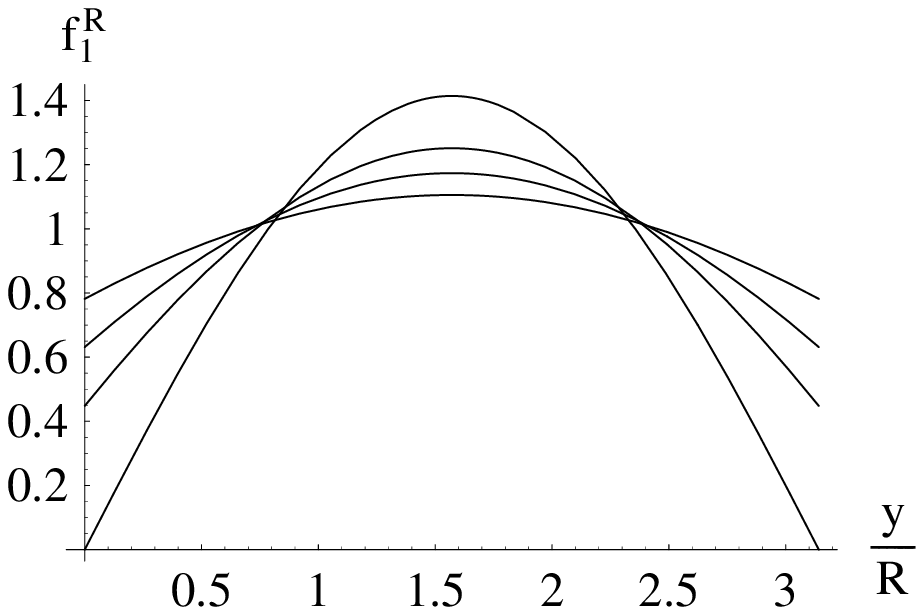,width=5cm} %\\
\caption{Wave function of the LH (left) and RH (right) components of the first
KK mode for different values of $a^L_0=a^L_\pi$. The different
lines correspond to $a_0/R=0,1,2,4$ for $f_1(0)$ from top to bottom
(on the left) for the LH component, and the opposite for the RH one.
\label{f1:ued}}
\end{center}
\end{figure}

\begin{figure}[!b]
\begin{center}
\epsfig{file=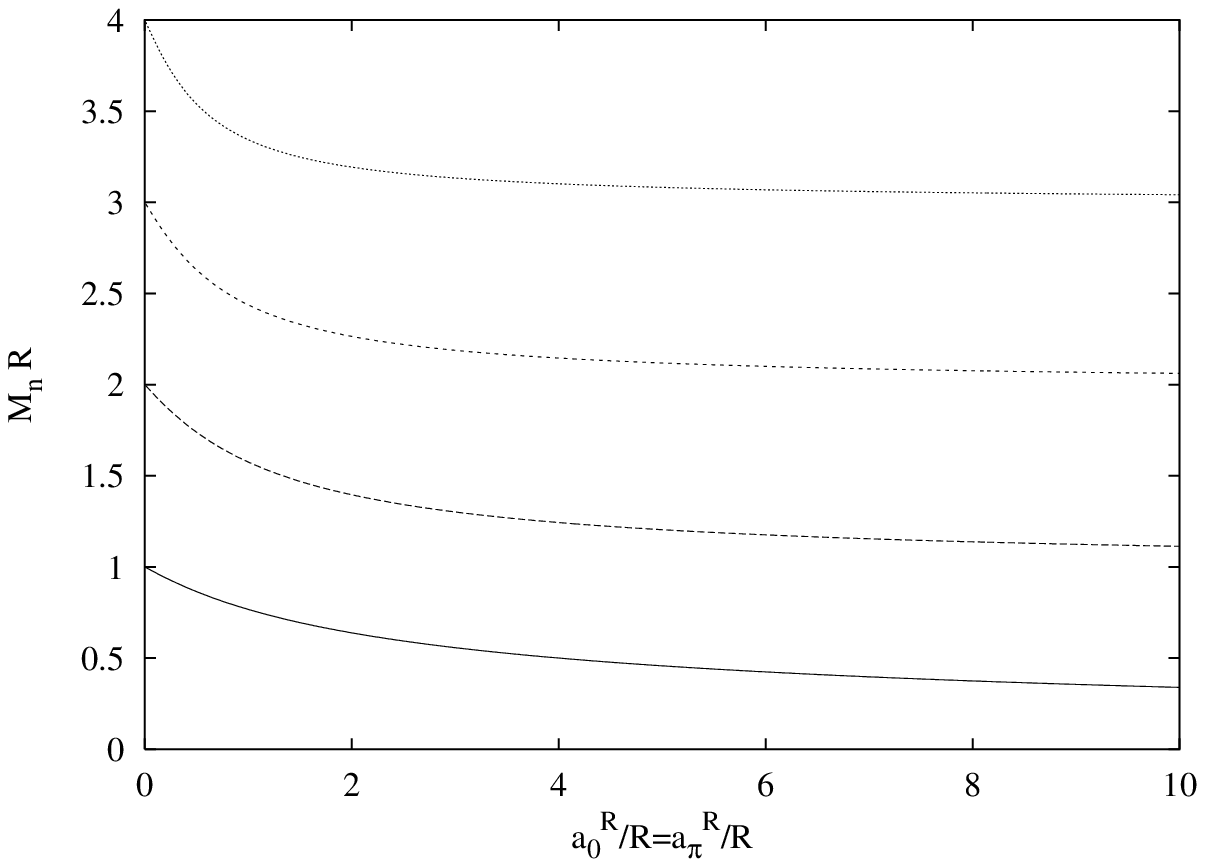,height=4cm}
\epsfig{file=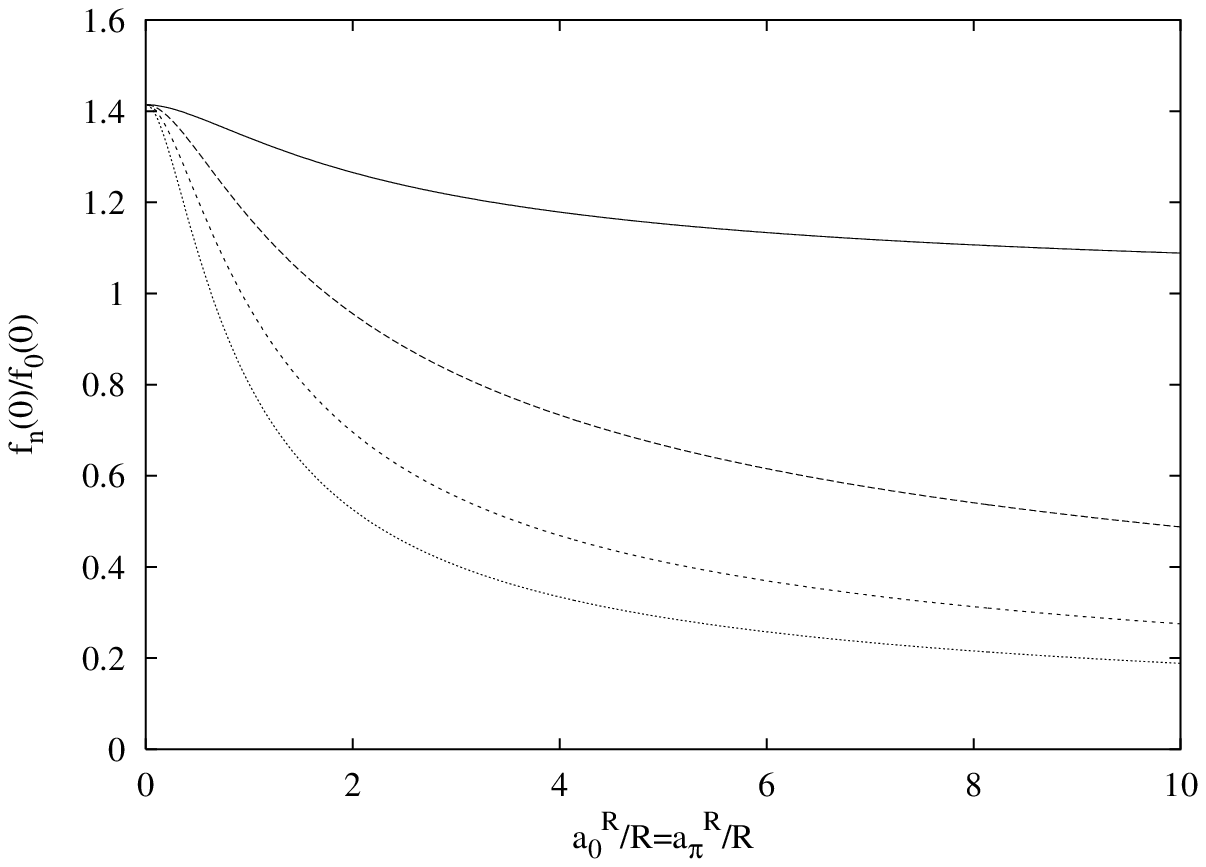,height=4cm} %\\
\caption{Masses (left) and LH couplings to the brane at $y=0$ normalized
to the zero mode coupling (right) for the first few KK modes
($n = 1,2,3,4$ from bottom to top (left) and from top to bottom
(right)) as a function of $a^L_0=a^L_\pi$. \label{spectrum:ued}}
\end{center}
\end{figure}

\begin{figure}[!t]
\begin{center}
\epsfig{file=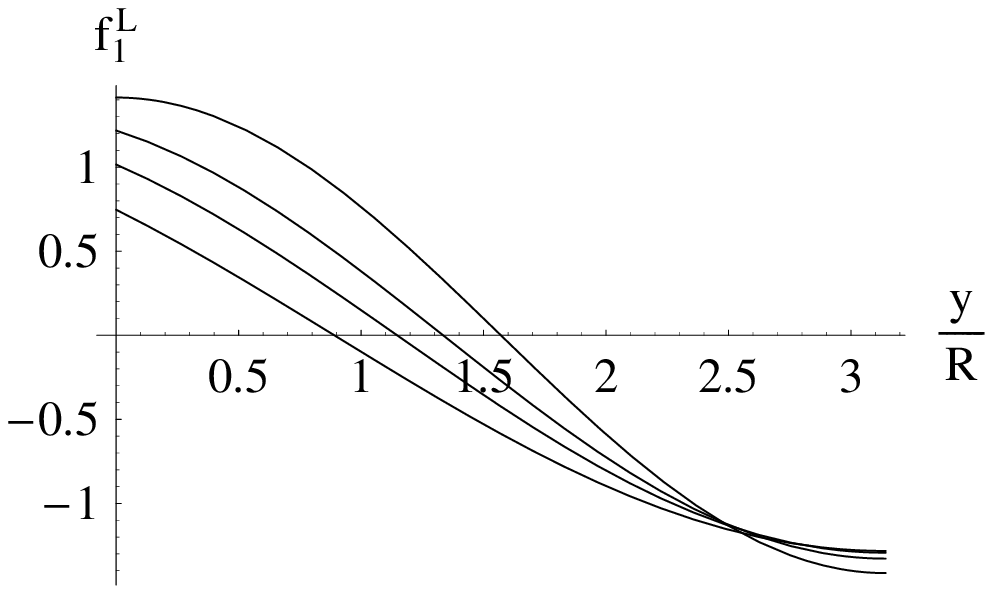,width=5cm}\hspace{.5cm}
\epsfig{file=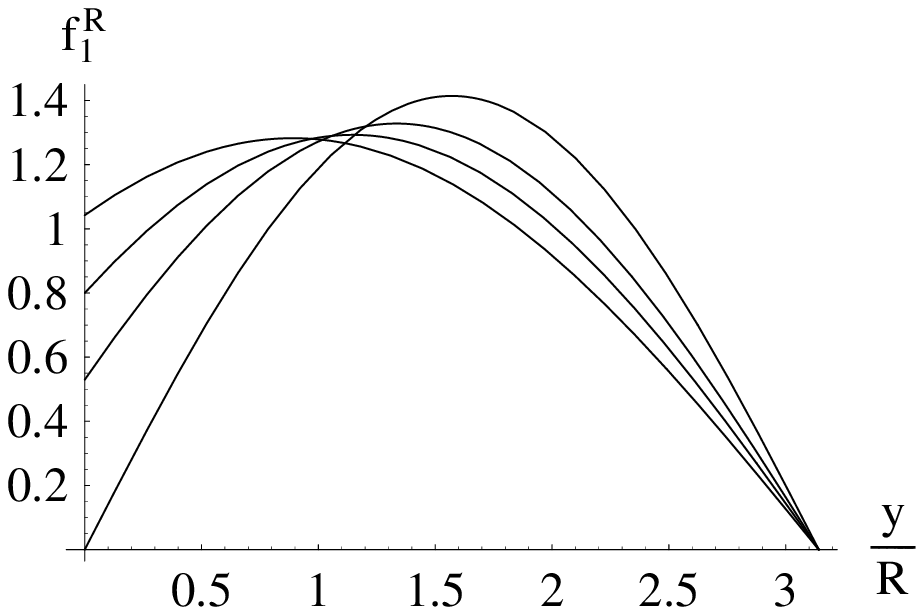,width=5cm} %\\
\caption{Wave function of the LH (left) and RH (right) components of the first
KK mode for $a^L_\pi=0$ and different values of $a^L_0$. The different
lines correspond to $a^L_0/R=0,1,2,4$ for $f_1(0)$ from top to bottom
(on the left) for the LH component, and the opposite for the RH one.
\label{f1:onebrane}}
\end{center}
\end{figure}

\begin{figure}[!b]
\begin{center}
\epsfig{file=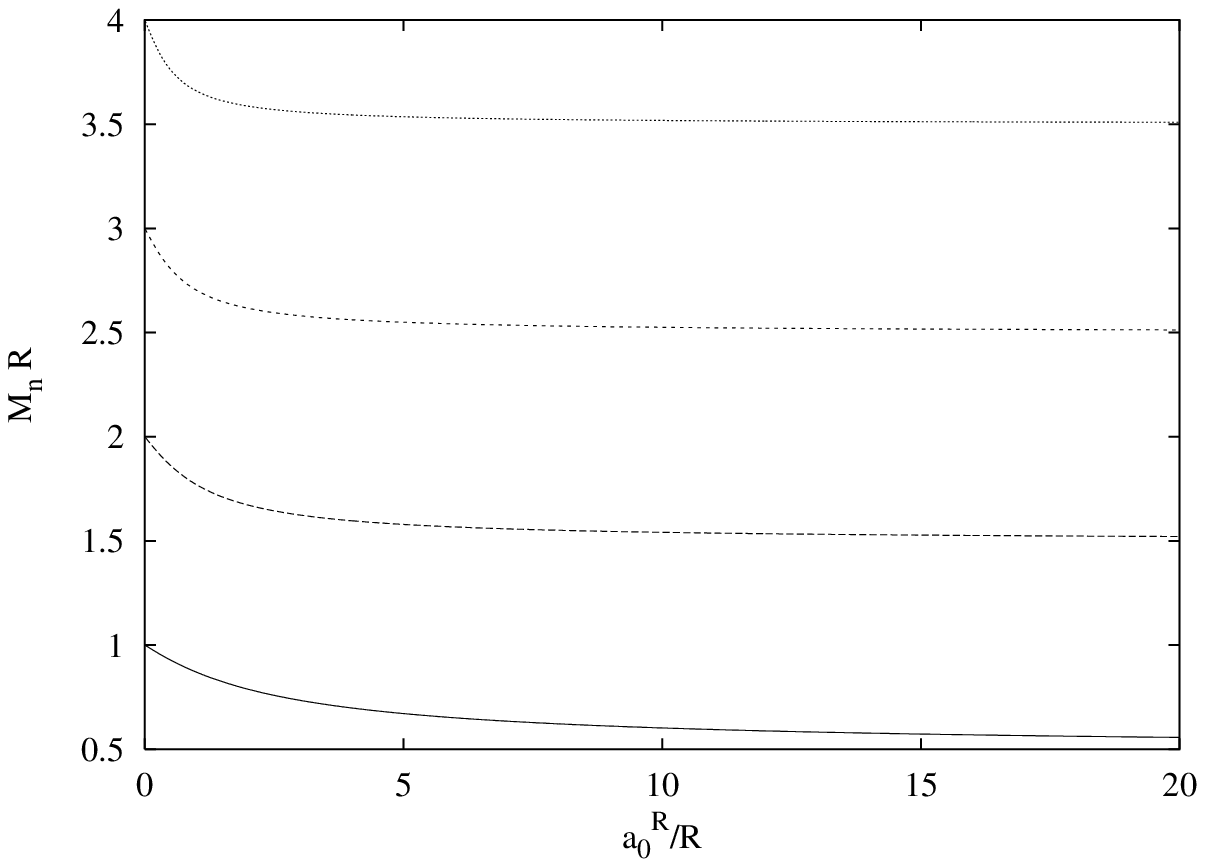,height=4cm}
\epsfig{file=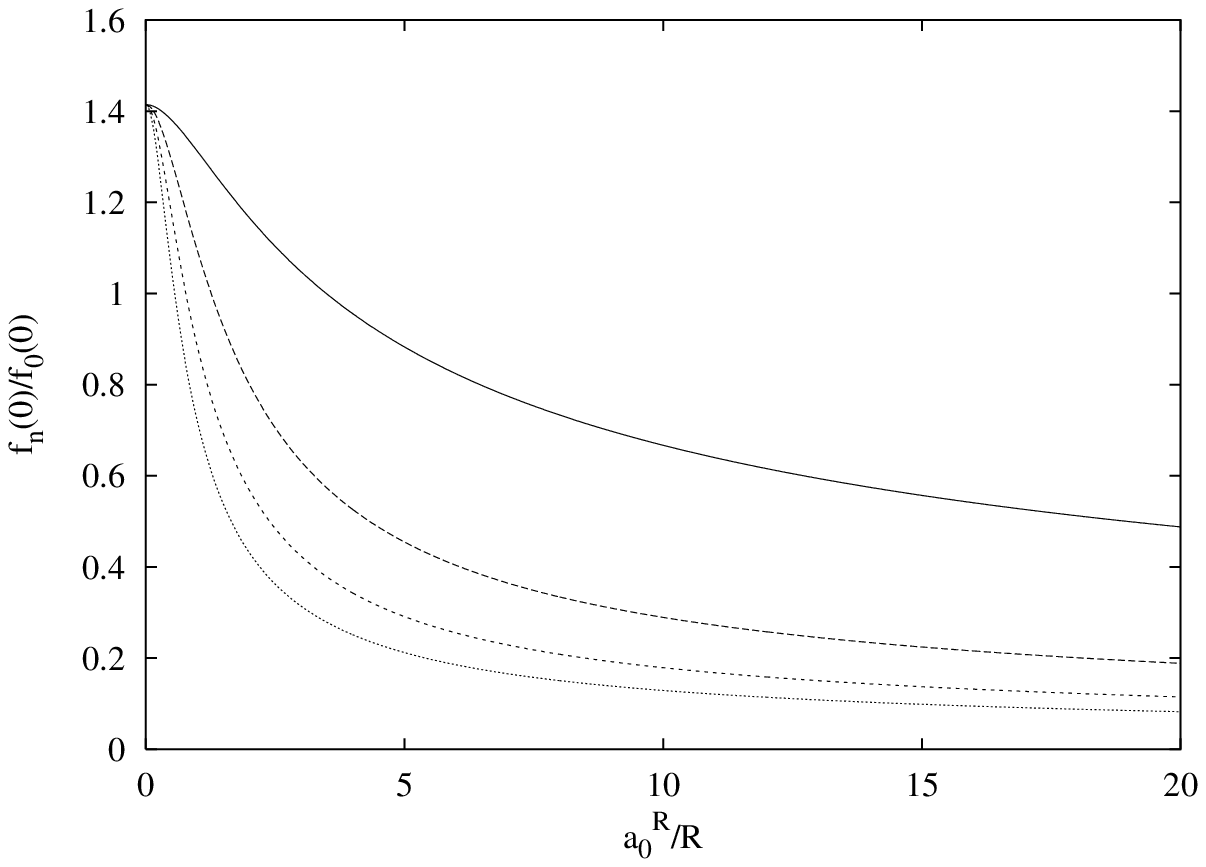,height=4cm} %\\
\caption{Masses (left) and LH couplings to the brane at $y=0$ normalized to
the zero mode coupling (right) for the first few KK modes
($n = 1,2,3,4$ from bottom to top (left) and from top to bottom
(right)) as a function of $a^L_0$, when $a^L_\pi=0$ .
\label{spectrum:onebrane}}
\end{center}
\end{figure}

Let us discuss fermions first. In five dimensions they 
are Dirac spinors with two chiral
components from the four-dimensional point of view:
$\Psi=\Psi_L+\Psi_R$, $\gamma^5 \Psi_{L,R}=\mp \Psi_{L,R}$.
Invariance
of the bulk kinetic term requires that the left-handed (LH) and
right-handed (RH) components have opposite $Z_2$ parities.
We choose the LH and RH components to be even and odd,
respectively. Taking into account possible BKT, the general kinetic
Lagrangian (with gauge couplings) for a fermion reads 
\begin{align}
\mathcal{L}=& 
\big(1+a^L_I\delta_I \big)
\bar{\psi}_L \mathrm{i} \not \! D \psi_L 
+ \big(1+a^R_I\delta_I \big) 
\bar{\psi}_R \mathrm{i} \not \! D \psi_R 
-\big(1+\frac{b_I}{2}\delta_I \big) \bar{\psi}_L D_y \psi_R 
\nonumber \\
&
- \frac{b_I}{2} \delta_I \big( D_y \bar{\psi}_R\big) \psi_L
+\big(1+\frac{c_I}{2} \delta_I \big) \bar{\psi}_R D_y \psi_L
+ \frac{c_I}{2} \delta_I \big( D_y \bar{\psi}_L\big) \psi_R
, \label{BKT:fermions}
\end{align}
where a sum over $I=0,\pi$ is understood and $\delta_I\equiv
\delta(y-I R)$. The BKT which contain $y$-derivatives, \textit{i.e.}, those 
proportional to $b_I$ and $c_I$, give rise to the
non-analytical behaviour we have mentioned in the
introduction, which has its origin in the fact that the branes in our
orbifold are lower-dimensional subspaces, \textit{i.e.}, they have zero
width. This situation may change at 
small distances in a string theory construction, as the brane gets
effectively a microscopic substructure. Within an effective
field-theoretical approach, a smooth, well-behaved theory can be
recovered implementing at the classical level a renormalization which
takes care of $\delta_0^2$-like 
singularities. We direct the
interested reader to Ref.~\cite{Aguila03a} for the details. Here we
simply disregard these BKT setting $b_{0,\pi}=c_{0,\pi}=0$. On the
other hand, the terms proportional to $a_I^R$ might naively be
argued to vanish, based on the odd character of $\Psi_R$. However,
this is not necessarily so if $\Psi_R$ is discontinuous at the branes,
and it turns out that this is the case when $a_I^R$ does not
vanish. Nevertheless, in the following we also take $a_I^R=0$, which
is stable under radiative corrections, to
reduce the number of independent parameters in our analysis.
Finally, we work with massless
fermions\footnote{In general, one could write Dirac mass terms with
masses which are odd functions of $y$ . Such masses may arise from the
vev of an odd scalar.}. 
%So, summarizing, we restrict our discussion to a theory with fermions
%described by the free Lagrangian
%\begin{eqnarray} 
%\mathcal{L}&=& 
%(1+a_I^L \delta_I) \bar{\Psi}_L \mathrm{i} \gamma^\mu D_\mu \Psi_L 
%%+(1+a^I_R \delta_I) 
%+\bar{\Psi}_R \mathrm{i} \gamma^\mu D_\mu \Psi_R 
%\nonumber \\ 
%&& 
%+\bar{\Psi}_R \partial_y \Psi_L-\bar{\Psi}_L \partial_y \Psi_R,
%\label{BKT:fermions}  
%\end{eqnarray} 
The spectrum in four dimensions can be computed by inserting the
Kaluza-Klein (KK) expansion of the field,
\begin{equation} 
\Psi_{L,R}(x,y)=\sum_{n=0}^\infty \frac{f_n^{L,R}(y)}{\sqrt{2\pi R}} 
\psi^{(n)}(x), 
\end{equation} 
into the action, and requiring the kinetic terms to be diagonal and
canonically normalized and the mass terms to be diagonal. This is
achieved by the following normalization and eigenvalue conditions:
\begin{equation} 
\int_{-\pi R}^{\pi R}\mathrm{d}y\; (1+a_0^L\delta_0 +a_\pi^L
\delta_\pi)  
\frac{(f_n^L)^2}{2\pi R} 
=\int_{-\pi R}^{\pi R}\mathrm{d}y\; 
\frac{(f_n^R)^2}{2\pi R} 
=1, \label{normalization} 
\end{equation} 
and 
\begin{eqnarray} 
\partial_y f_n^R&=& m_n(1+a^L_0 \delta_0+a^L_\pi \delta_\pi) f_n^L,
\label{eigen:LH}\\  
\partial_y f_n^L&=&-m_n f_n^R.\label{eigen:RH} 
\end{eqnarray} 
The eigenvalue equations can be solved by iteration, which leads to
a quadratic equation for the even component,
\begin{equation} 
\partial_y^2 f_n^L=- m_n^2(1+a^L_0 \delta_0+a^L_\pi \delta_\pi) f_n^L, 
\end{equation} 
while the odd component can be obtained directly from Eq.(\ref{eigen:RH}). 
The result is a massless zero mode only for the even component (therefore 
chirality is recovered thanks to the orbifold projection) with a flat 
wave function,
\begin{equation} 
f_0^L=\frac{1}{\sqrt{1+\frac{a^L_0+a^L_\pi}{2\pi R}}}, 
\end{equation} 
plus a tower of vector-like massive KK modes with oscillating wave
functions  
\begin{equation} 
f_n^L(y)=A_n [\cos(m_n y)-\frac{a^L_0 m_n}{2} \sin (m_n y)], 
\end{equation} 
and 
\begin{equation} 
f_n^R(y)=A_n [\sin(m_n y)+\frac{a^L_0 m_n}{2} \cos (m_n y)]. 
\end{equation} 
$A_n$ is a normalization constant determined by Eq. 
(\ref{normalization}). Notice that the BKT make
the odd component discontinuous across the branes. 
The masses are determined by the following 
equation 
\begin{equation} 
(4-a^L_0 a^L_\pi m_n^2) \tan (m_n \pi R)+2(a^L_0+a^L_\pi)m_n=0. 
\end{equation} 
As anticipated in the introduction, the presence of BKT modifies
the spectrum. The modifications depend to a great extent
on whether the sizes of the BKT at both fixed 
points are comparable or not. If they are, the first mode can be made 
arbitrarily light, with mass 
\begin{equation} 
m_1^2\sim 2 \frac{a^L_0+a^L_\pi}{a^L_0 a^L_\pi \pi R}, 
%\quad
%(a^L_{0,\pi} \gg R),  
\end{equation} 
and couplings to brane fields equal in size to the one of 
the zero mode for $a^L_{0,\pi} \gg R$, whereas the rest of the massive 
modes have masses which approach $m_n \sim (n-1)/R$ and decouple from the 
branes in that limit. The wave function of the LH and RH components of 
the first KK mode and the masses and relative couplings to the brane of the 
first four KK excitations of the even component are represented, respectively 
in Figs.~\ref{f1:ued} and~\ref{spectrum:ued} for the symmetric case
$a^L_0=a^L_\pi$.  
When one of the two BKT is negligible with respect to 
the other, all KK modes behave similarly, with masses approaching 
$m_n \sim (n-1/2)/R$ and couplings to the corresponding brane
increasingly small for large values of the BKT. In
Figs.~\ref{f1:onebrane} and~\ref{spectrum:onebrane}, we
represent the wave function of the first mode and the masses and
couplings to the brane of the first four KK excitations of the even
component as a function of $a^L_0$ for $a^L_\pi=0$.

%%%%%%%%%%%%%%%%%%%%%%%%%%%%%%%%%%%%%%%%%%%%%%%%%%%%%%%%%%%%%%%%%
%%%%%%%%%%%%%%%%%%%%%%%%%%%%%%%%%%%%%%%%%%%%%%%%%%%%%%%%%%%%%%%%%
%%%%%%%%%%%%%%%%%%%%%%%%%%%%%%%%%%%%%%%%%%%%%%%%%%%%%%%%%%%%%%%%%
%%%%%%%%%%%%%%%%%%%%%%%%%%%%%%%%%%%%%%%%%%%%%%%%%%%%%%%%%%%%%%%%%
%%%%%%%%%%%%%%%%%%%%%%%%%%%%%%%%%%%%%%%%%%%%%%%%%%%%%%%%%%%%%%%%%

Let us now discuss the situation for gauge
bosons in the $S^1/Z_2$ orbifold~\cite{Carena02}. The fifth component
of a vector 
boson is forced to have $Z_2$ parity opposite to the one of the other
components. Unless one is interested in 
breaking (part of) the gauge group, the first four components must be
taken to be even. Breaking the gauge group by orbifold
projections is a very interesting possibility which has been exploited
in GUT models~\cite{GUTS}, but here we stick to even $A_\mu$ for
definiteness. The most general kinetic Lagrangian is
\begin{equation}
\mathcal{L} = -\frac{1}{4}
(1+a^A_I\delta_I)  \mathrm{tr} F_{\mu\nu}F^{\mu\nu} - \frac{1}{2}
(1+c_I^A\delta_I) \mathrm{tr} F_{5\nu}F^{5\nu} \,
. \label{eq:gaugebosonLag}  
\end{equation} 
In this case, gauge invariance forbids the most singular ``$b$-like''
BKT. The terms proportional to $c^A_I$ have no effect at all when
treated in a non-perturbative way, but give rise to singularities at
high orders of perturbation theory, which may be substructed by
(classical) counterterms~\cite{Aguila03a}. In this second approach, a
non-vanishing finite effect survives. Here, we simply set
$c^A_{0,\pi}=0$ and focus on arbitrary coefficients $a_I^A$.
%
%relevant Lagrangian is
%\begin{equation} 
%\mathcal{L}=-\frac{1}{2} \big[ \mathrm{Tr} F_{MN} F^{MN}+a_I^A
%\delta_I \mathrm{\Tr} F_{\mu\nu}F^{\mu\nu} \big], 
%\end{equation} 
The gauge invariant KK decomposition of~(\ref{eq:gaugebosonLag}) has
been carried out in~\cite{Aguila03a}. Working instead in the ``axial''
gauge $A_5=0$ and expanding in KK modes the even
components,
\begin{equation} 
A_\mu(x,y)=\sum_n \frac{f^A_n(y)}{\sqrt{2\pi R}} A_\mu^{(n)}(x), 
\end{equation} 
we arrive at the same eigenvalue and orthonormality equations 
as the ones we have obtained above for the even component of the
bulk fermion when only $a_{0,\pi}^L$ are
non-vanishing~\cite{Carena02}. The spectrum and wave functions are
therefore identical to the one we have just discussed for (even)
fermions.

%%%%%%%%%%%%%%%%%%%%%%%%%%%%%%%%%%%%%%%%%%%%%%%%%%%%%%%%%%%%%

\section{Phenomenology of brane kinetic terms}
We have seen in the previous section how the presence of BKT
affects the masses and wave functions of the KK modes of bulk fermions
and bosons. To extract phenomenological implications, we need to
compute the couplings of the four-dimensional effective theory, which
are given by the overlap of wave functions (with are delta functions for
fields on the branes) times the five-dimensional couplings. 
The impact of BKT on Yukawa couplings can be very relevant
phenomenologically, as has been shown in~\cite{Aguila03b}. 
Nonetheless we will mainly concentrate on gauge couplings, as they
affect directly the very precisely measured electroweak observables, 
and have not been previously studied when both bulk gauge bosons
and bulk fermions have non-vanishing BKT.

%\subsection{Gauge interactions}

The gauge interactions of fermions can be obtained from the KK
reduction of the fermionic Lagrangian, Eq.(\ref{BKT:fermions}). They
read 
\begin{equation}
\mathcal{L}_{\mathrm{int}}=- \sum_{mnr} g_{mnr} \bar{\psi}_L^{(m)}
\gamma^\mu A_\mu^{(r)} 
\psi_L^{(n)},
\end{equation}
with effective gauge couplings given by
\begin{equation}
g_{mnr}=\frac{g_5}{\sqrt{2\pi R}} \int_{-\pi R}^{\pi R}
\mathrm{d}y\;  (1+a_I^L\delta_I) \frac{f_m^L f_n^L f^A_r}{2\pi R}.
\end{equation}
The phenomenologically relevant couplings are the ones of two fermion
zero modes to the KK excitations of gauge bosons, 
\begin{equation}
\frac{g^{(00n)}}{g^{(000)}}=
\frac{\sqrt{1+\frac{a_0^A+a_\pi^A}{2\pi R}}}{1+\frac{a_0^L+a_\pi^L}{2\pi R}}
\Big[ \frac{a_0^L-a_0^A}{2\pi R} f_n^A(0)
+\frac{a_\pi^L-a_\pi^A}{2\pi R} f_n^A(\pi R)\Big].\label{gn/g0}
\end{equation}
Note that these couplings vanish when $a_0^L=a_0^A$ and
$a_\pi^L=a_\pi^A$, and in particular in the limit of no BKT
(due to KK number conservation).

\begin{figure}[h]
\begin{center}
\epsfig{file=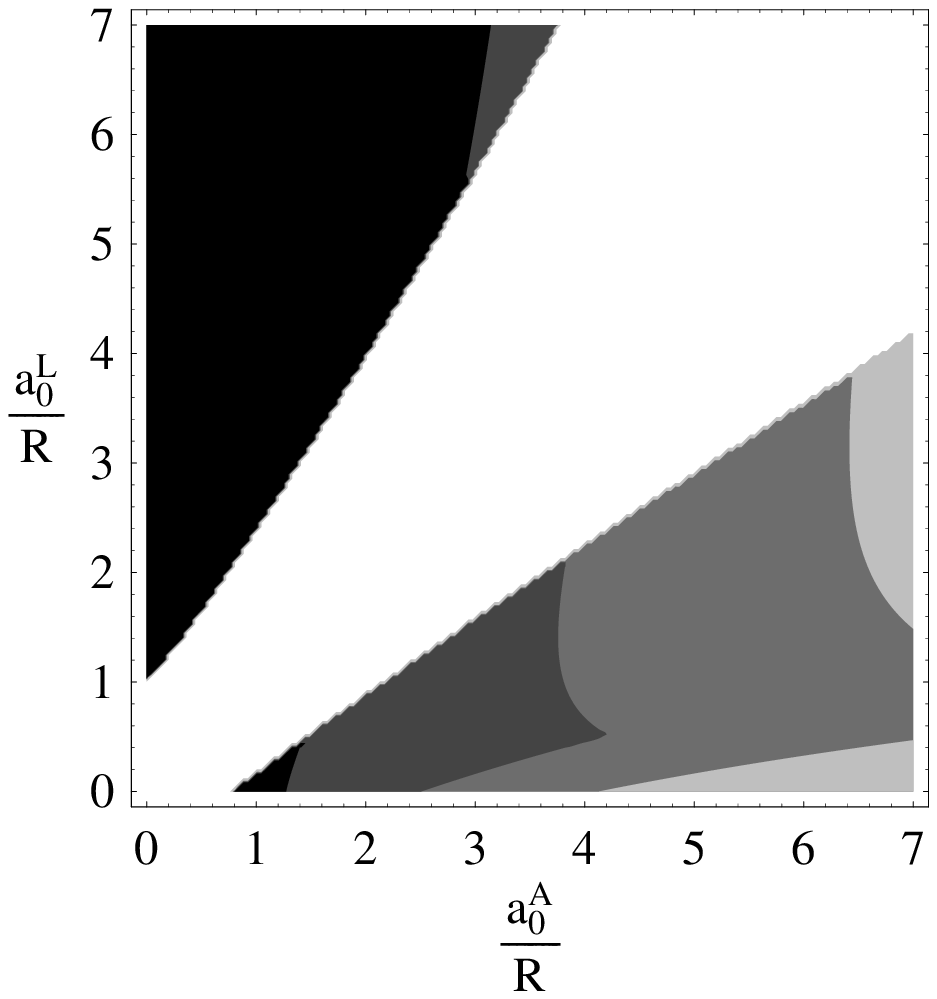,width=5cm}\hspace{.5cm}
\epsfig{file=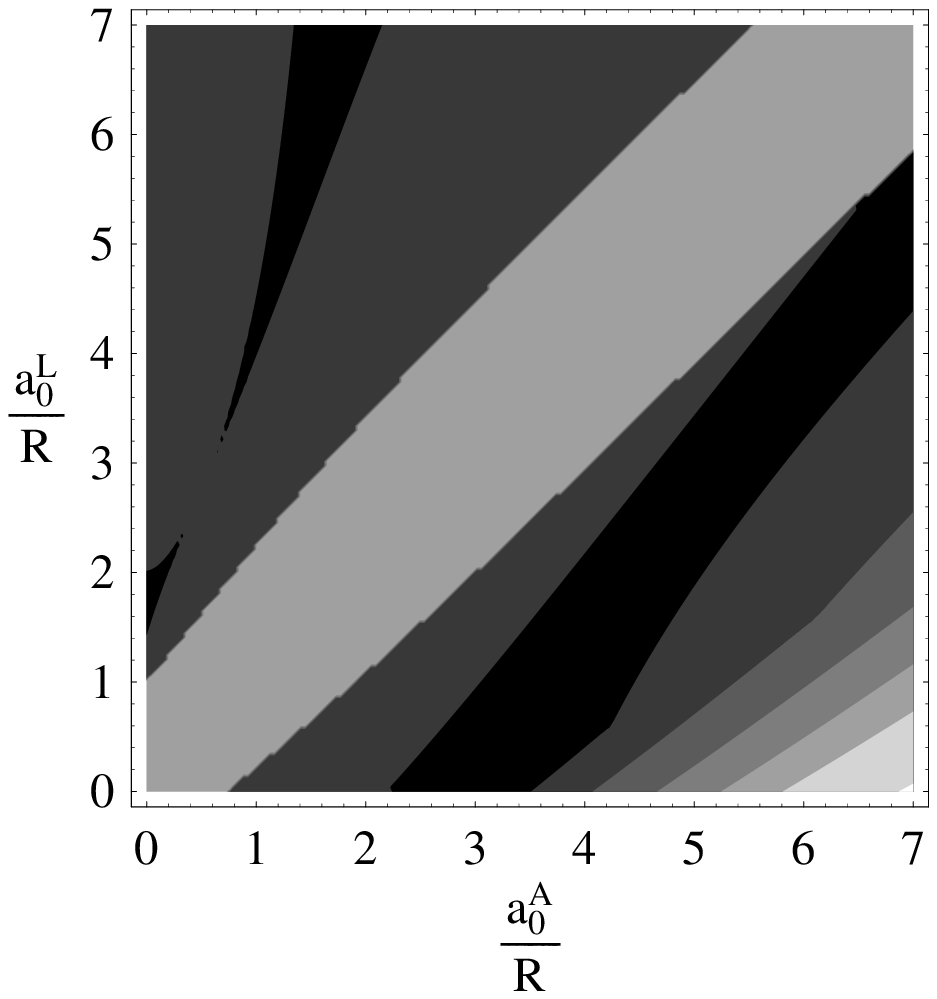,width=5cm} %\\
\caption{Bound on the compactification scale as a function of
$a^A$ (horizontal axis) and $a^L$ (vertical axis) for BKT equal at
both branes (left) and only at $y=0$ (right) respectively. The
band along the diagonal $a^A=a^L$ corresponds to $U_\mathrm{eff}$ being
too small to give a meaningful lower bound. The rest of the bands
correspond, from 
dark to light, to $M_c\geq 2,3,3.5,4\;\mathrm{TeV}$ (left) and
$M_c\geq 1,2,3,4,5,6 
\;\mathrm{TeV}$ (right).
\label{Mcbound:fig}}
\end{center}
\end{figure}

A fit to electroweak precision observables taking into account the
modifications induced by the presence of the gauge boson KK modes
results in bounds on the compactification scale, as a function of the
fermion and gauge boson BKT. We adapt
the methods of Ref.~\cite{Carena03} to our particular
case. Considering all gauge bosons to share the same BKT on each
brane, with common coefficient $a^A_I$, and 
similarly for fermions, with coefficient $a^L_I$, but allowing for
independent $a^A_I$ and $a^L_I$, our model is
described by the following parameters:
%\begin{equation}
$R,a^A_I,a^L_I,g_5,g_5^\prime$ and $v$,
%\end{equation}
where $g_5,g_5^\prime$ are the gauge couplings of $\mathrm{SU}(2)_L$
and $\mathrm{U}(1)_Y$, respectively, and $v$ is the vev of the Higgs
boson, which we consider localized at $y=0$. The idea
is to compute the corrections due to the extra dimensions to all
observables, fix three of our independent parameters in terms of three
observables and perform the fit as a function of the remaining
parameters. We fix $g_5,g_5^\prime$ and $v$ using
$\alpha(M_z)\simeq 1/129$, $M_Z\simeq 91.2 \,\mathrm{GeV}$ and
$G_F\simeq 1 \times 10^{-5} \, \mathrm{GeV}^{-2}$. All the observables
are then expressed in
terms of them plus the compactification scale and the BKT. There are
two types of modifications due to the presence of extra dimensions, one is
the usual four-fermion interactions mediated by gauge boson KK
excitations, the other is due to the localized Higgs and can be traced
to the different mixing of the $W$ and $Z$ with their
respective KK towers or, alternatively, to the fact that if the KK
expansion is 
performed including the localized Higgs vev, the $Z$ and $W$ zero
modes are no longer flat. All these modifications can be
captured in the definition of effective oblique parameters
$T_\mathrm{eff}, S_\mathrm{eff}, U_\mathrm{eff}$ (we call
them effective because they actually contain the leading
non-oblique effects as well). The
localized Higgs effect only affects $T_\mathrm{eff}$ whereas the others
depend on the couplings in Eq.(\ref{gn/g0}) and therefore vanish in
the limit $a^L_I=a^A_I$. The KK contributions to $S, T, U$ are
determined by matching the effective Lagrangian for zero modes, with
fermion interactions rescaled to unity, to the following generic
Lagrangian with oblique corrections:
\begin{eqnarray}
\mathcal{L}&=&
-\frac{1-\Pi^\prime_{WW}}{2g^2} W^+_{\mu\nu} W^{\mu\nu}_-
-\frac{1-\Pi^\prime_{ZZ}}{4(g^2+g^{\prime\;2})}
Z_{\mu\nu}Z^{\mu\nu}
\\
&-&
\bigg(\frac{1}{4\sqrt{2}G_F}+\frac{\Pi_{WW}(0)}{g^2}\bigg)
W^+_\mu W^\mu_-
-\frac{1}{2}\bigg(\frac{1}{4\sqrt{2}G_F}
+\frac{\Pi_{ZZ}(0)}{g^2+g^{\prime\;2}}
\bigg)
Z_\mu Z^\mu,
\nonumber
\end{eqnarray}
where $g\equiv g_5 f_0^A/\sqrt{2\pi R}$,
$g^\prime\equiv g^\prime_5 f_0^A/\sqrt{2\pi R}$ and
we have only represented the relevant terms. The
oblique parameters are defined in terms of the ``self-energies''
above (including non-standard tree-level contributions) as
\begin{eqnarray}
&S=&
16 \pi
\bigg(\frac{\Pi^\prime_{ZZ}}{g^2+g^{\prime\;2}}
-\Pi^\prime_{3Q}\bigg), \\
&T=&
\frac{4\pi}{s^2c^2m_Z^2}
\bigg(\frac{\Pi_{WW}(0)}{g^2}
-\frac{\Pi_{ZZ}(0)}{g^2+g^{\prime\;2}}
\bigg), \\
&U=&
16 \pi
\bigg(\frac{\Pi^\prime_{WW}}{g^2}
+
\frac{\Pi^\prime_{ZZ}}{g^2+g^{\prime\;2}}
\bigg),
\end{eqnarray}
where $\Pi^\prime_{3Q}$ is related to the $Z-\gamma$ kinetic mixing
which is zero in our case and 
$s\equiv g_5^\prime/\sqrt{g_5^2+g_5^{\prime\;2}}=\sin\theta_W
+\mathcal{O}(v^2/m_n^2)$ and
$c\equiv g_5/\sqrt{g_5^2+g_5^{\prime\;2}}=
\cos\theta_W+\mathcal{O}(v^2/m_n^2)$, with $\theta_W$ the Weinberg
angle.
We also have to take into account non-oblique four fermion
interactions, which can be parameterized by
\begin{equation}
\Delta T=-\frac{1}{\alpha}\frac{\delta G_F}{G_F}=-\frac{1}{\alpha}
\sum_n \bigg(\frac{g^{(00n)}}{g^{(000)}}\frac{m_W}{m_n}\bigg)^2.
\end{equation}
In order to include this non-oblique effect in the electroweak fit
we redefine $T\to T+\Delta T$ and $U\to
U-4 s^2 \Delta T$~\cite{Carena03}. 
The tree-level contribution of extra dimensional
physics to the resulting effective oblique parameters reads
\begin{eqnarray}
\bar{S}_\mathrm{eff}&=&\bar{S}=-\frac{8s^2c^2}{\alpha} \sum_n
\frac{g^{(00n)}}{g^{(000)}}\frac{f_n^A(0)}{f_0^A(0)} \frac{m_Z^2}{m_n^2},
\\
\bar{T}_\mathrm{eff}&=&
\bar{T}+\Delta T=\frac{1}{\alpha} \sum_n\Bigg[
\bigg(
\frac{f_n^A(0)}{f_0^A(0)}
-2\frac{g^{(00n)}}{g^{(000)}}
\bigg)
\frac{f_n^A(0)}{f_0^A(0)}
\frac{m_Z^2-m_W^2}{m_n^2}
\nonumber \\
&&
\phantom{T+\Delta T=\frac{1}{\alpha} \sum_n\Bigg[
%\bigg(
}-\bigg(\frac{g^{(00n)}}{g^{(000)}}\bigg)^2
\frac{m_W^2}{m_n^2}
\Bigg],
\\
\bar{U}_\mathrm{eff}&=&\bar{U}-4s^2\Delta T=
\frac{4s^2}{\alpha} \sum_n
\bigg(\frac{g^{(00n)}}{g^{(000)}}\bigg)^2
\frac{m_Z^2}{m_n^2},
\end{eqnarray}
where the bars indicate that only tree level contributions are
included.
Note that, being already of order $v^2/m_n^2$, we can plug the
experimental results for $s,c,m_Z$ and $m_W$ in the above expressions.
A global fit to the electroweak observables
gives~\cite{Erler}
%, for the new physics contribution to the oblique parameters 
%\cite{Altarelli:2001wx}
\begin{eqnarray}
S_\mathrm{new}&=&-0.03\pm 0.11, \\
T_\mathrm{new}&=&-0.02\pm0.13, \\
U_\mathrm{new}&=&0.24\pm0.13,
\end{eqnarray}
for $m_t=173$~GeV and $m_H=115$~GeV and  ``new'' here stands for beyond the
SM contribution.
%The subscript New indicates that the SM contribution has been
%already subtracted.
% and therefore they are directly applicable to our effective
%parameters. 
%We consider here a light Higgs so that the results of this fit are
%still valid. 
Requiring that the contribution to each of the three
effective oblique parameters remains within the 1-$\sigma$ interval,
we obtain the bounds on the compactification scale shown in 
Fig.~\ref{Mcbound:fig} as a function of the gauge boson
(horizontal axis) and fermion (vertical axis) BKT, for BKT equal
at both branes and for BKT only at $y=0$. In both cases there is
a band along the diagonal $a^L=a^A$ where $U_\mathrm{eff}$ is too
small to give a meaningful bound on the compactification scale. The
rest of the bands correspond to
$M_c\geq 2,3,3.5,4\;\mathrm{TeV}$ for the case that $a^L_0=a^L_\pi$,
$a^A_0=a^A_\pi$ (left plot) and $M_c\geq 1,2,3,4,5,6
\;\mathrm{TeV}$ when $a_\pi^L=a_\pi^A=0$ (right plot).
Note that in the symmetric case the bounds are less strict, even
though in that case the first KK excitation is lighter. The reason
is that the first KK mode decouples when $a_0^L=a_\pi^L$ (see
Eq. (\ref{gn/g0}) and note that $f^A_1(\pi R)=-f^A_1(0)$ when
$a^A_0=a^A_\pi$). Therefore,
only a suppressed contribution to $T_\mathrm{eff} \sim (m_Z^2-m_W^2)/m_1^2$
effectively bounds $M_c$. In the case of BKT just at one brane
%, on the
%other hand, 
the added effect of all modes is much stronger. It is also
apparent from the figures that the dependence on the fermion and gauge
boson BKT is highly asymmetric in this case, with a stronger
dependence on the gauge boson BKT than on the fermion ones. Finally,
we observe that the effect of decoupling from the brane with BKT,
which we pointed out in Section~\ref{sectionKK}, is not very
important, as far as gauge interactions are concerned, in models in
which both gauge bosons and fermions live in the bulk.

%%%%%%%%%%%%%%%%%%%%%%%%%%%%%%%%%%%%%%%%%%%%%%%%%%%%%%%%%%%%%%%%%%
%%%%%%%%%%%%%%%%%%%%%%%%%%%%%%%%%%%%%%%%%%%%%%%%%%%%%%%%%%%%%%

As we mentioned at the beginning of this section, Yukawa couplings are
also modified by the presence of BKT for fermions. Models with bulk
fermions have a non-unitary CKM matrix due to the mixing of zero and
massive KK modes~\cite{delAguila:2000aa}. 
In many models the Higgs lives on one of the
branes, say the one at $y=0$, and then the departure from the SM is
proportional to the mass of the quarks involved, i.e., it is most
relevant for top physics~\cite{delAguila:2000kb}. 
These effects give rise to stringent
limits from the T parameter, which become weaker when
$a_0\gg a_\pi$. Details can be found in~\cite{Aguila03b}.

\section{Conclusions}
We have seen that the BKT for bulk fields have a significant impact on
the phenomenology of compact extra dimensional theories when the
corresponding coefficients are comparable to or larger than the
compactification scale. This means that the parameter space of the
theory without BKT is enlarged, and this allows for a greater freedom
in the construction of models which are consistent with present
data (and possibly with implications observable in future
experiments). Of course, the price to pay is a reduction of
predictivity in a general effective approach. 
\newline

This work has been supported in part by MCYT under contract
FPA2000-1558, by Junta de Andaluc\'{\i}a group FQM 101, by the
European Community's Human Potential Programme under contract
HPRN-CT-2000-00149 Physics at Colliders, and by PPARC.

\end{document}